\documentstyle[preprint,epsf,aps]{revtex}
\tighten
\begin{document}

\draft
\title{Spin correlations in nonlinear optical response:
Light-induced Kondo effect}
\author{T. V. Shahbazyan and I. E. Perakis}
\address{Department of Physics and Astronomy,  
Vanderbilt University, Nashville, TN 37235}
\author{M. E. Raikh}
\address{Department of Physics, University of Utah, Salt Lake City,
UT 84112}
\maketitle
\begin{abstract}    
We study  the role of spin correlations in 
nonlinear  absorption due to optical transitions from a
deep impurity level to states above a Fermi sea. We demonstrate that
the Hubbard repulsion between two electrons at the
impurity leads to a logarithmic divergence in the third-order
optical susceptibility $\chi^{(3)}$ at the absorption
threshold. This divergence is a manifestation of the Kondo physics in
the nonlinear optical response of  Fermi sea systems. 
We  also show that, for off-resonant pump excitation, 
the pump-probe spectrum exhibits  a narrow peak below 
the linear absorption onset.
Remarkably, the light-induced  Kondo temperature, which governs 
the shape of the Kondo-absorption spectrum, can be {\em tuned} 
by varying the intensity and frequency of the pump. 
\end{abstract}

\narrowtext
There are two prominent many-body effects 
in the linear  absorption
spectrum due to optical transitions
from a localized impurity level 
to the continuum of states above a Fermi
sea (FS). First is the Mahan 
singularity due to the attractive
interaction between the FS and the localized hole. 
Second is the Anderson orthogonality
catastrophe due to the readjustment 
of the FS density profile during the optical 
transition. Both  effects have long become textbook
material\cite{mahan-book}. 
The role of many-body correlations in the {\em nonlinear} optical
response has only been investigated
during the last decade \cite{reviews}.
Recently, there has been a growing interest in the coherent
ultrafast dynamics of the FS systems at low
temperatures\cite{bre95,portengen,awschalom,hall,dodge99,FES}.

In this paper, we suggest a new many-body effect in the nonlinear
absorption of a FS system with a deep impurity level. 
This effect originates from the {\em spin} correlations between 
the photoexcited and the FS electrons.
We note that a number of different 
intermediate processes contribute to the third--order 
optical susceptibility $\chi^{(3)}$\cite{mukamel-book}. 
It is crucial that, in the system under study,
some of the intermediate states involve a doubly-occupied impurity
level. For example, the optical field can first cause a transition 
of a FS electron  to the singly-occupied impurity level, 
which thus becomes doubly-occupied, and then excite both electrons
from the impurity level to the conduction band. 
This is illustrated in Fig.\ 1(a). Important is that, while
on the impurity, the two electrons experience a Hubbard repulsion. 
Our main observation is that such a repulsion gives rise to an anomaly in
$\chi^{(3)}$. The origin of this anomaly is intimately related to the
Kondo effect.

To be specific, we restrict ourselves to 
pump-probe spectroscopy, where 
a strong pump and a weak probe optical field
are applied to the system and  the optical 
polarization along the probe direction is measured. 
We only consider near-threshold absorption at zero temperature
and assume that the pump frequency is tuned below 
the onset of optical transitions from the impurity 
level so that dephasing processes due to electron-electron and
electron-phonon interactions are suppressed. 
Under such excitation conditions, the following 
Hamiltonian describes the system:
$H_{tot}=H+H_{1}(t)+H_{2}(t)$, where 
\begin{equation}
\label{Ham}
H=\sum_{{\bf k}\sigma}\varepsilon_kc_{{\bf k}\sigma}^{\dag}c_{{\bf k}\sigma}
+\varepsilon_d\sum_{\sigma}d_{\sigma}^{\dag}d_{\sigma}
+\frac{U}{2}\sum_{\sigma\neq\sigma'}\hat{n}_{\sigma}\hat{n}_{\sigma'},
\end{equation}
is the Hamiltonian in the absence of optical fields; here 
$c_{{\bf k}\sigma}^{\dag}$ and $d_{\sigma}^{\dag}$ are conduction 
and localized electron creation operators, respectively,
($\hat{n}_{\sigma}=d_{\sigma}^{\dag}d_{\sigma}$);
$\varepsilon_k$ and $\varepsilon_d$ are the corresponding energies,
and $U$ is the Hubbard interaction (all energies are measured from the
Fermi level). 
The coupling to the optical fields is described by the Hamiltonian 
$H_i(t)=-M_i(t)\hat{T}^{\dag} + h.c.$
where $\hat{T}^{\dag}=
\sum_{{\bf k}\sigma}c_{{\bf k}\sigma}^{\dag}d_{\sigma}$,
($i=1,2$ denotes the probe and pump, respectively) with
$M_i(t)=e^{i{\bf k}_i\cdot{\bf r}-i\omega_it}\mu{\cal E}_i(t)$.
Here ${\cal E}_i(t)$, ${\bf k}_i$, and $\omega_i$ are the 
pump/probe electric field amplitude, direction and 
central frequency, respectively, and $\mu$ is the dipole
matrix element.
The pump-probe polarization is obtained
by expanding the optical polarization, $\mu\,\langle\hat{T}\rangle$, to
the first order in $H_1$ and keeping the terms propagating in the
probe direction\cite{mukamel-book}:
\begin{eqnarray}\label{PP}
P(t)=
i\mu \int_{-\infty}^{t} dt' 
M_1(t') 
\biggl[\langle\Phi(t)|\hat{T}{\cal K}(t,t')
\hat{T}^{\dag} |\Phi(t')\rangle
-\langle\Phi(t')|\hat{T}^{\dag} {\cal K}(t',t)
\hat{T}|\Phi(t)\rangle\biggr],
\end{eqnarray}
where ${\cal K}(t,t')$ is the evolution operator for the Hamiltonian
$H+H_2(t)$ and the state 
$|\Phi(t)\rangle$  satisfies  the Schr\"{o}dinger equation 
$i\partial_t |\Phi(t)\rangle=[H+H_2(t)]|\Phi(t)\rangle$. 

The third order polarization is obtained by expanding  
${\cal K}(t,t')$ and $|\Phi(t)\rangle$ 
up to the second order in $H_2$. Below we consider sufficiently large
values of $U$ so that, in the absence of optical fields, the ground
state of $H$, $|\Omega_0\rangle$,    
represents a {\em singly-occupied} impurity  and  full FS.
For large $U$, the doubly-occupied impurity states are energetically
unfavorable and can be excluded from the expansion of the
polarization (\ref{PP}) with respect to $H_2$. 
The third-order pump-probe polarization then
takes the form 
$P^{(3)}(t)=e^{i{\bf k}_1 \cdot{\bf r}-i\omega_1t}\tilde{P}^{(3)}$ 
with
\begin{eqnarray}\label{PP3}
\tilde{P}^{(3)}=
i\mu^4\int_{-\infty}^{t}
dt' {\cal E}_1(t')e^{i\omega_1(t-t')}
\biggl[Q_1(t,t')
+Q_1^{*}(t',t)
+Q_2(t,t')+Q_3(t,t')\biggr],
\end{eqnarray}
where
\begin{eqnarray}\label{Q}
&&
Q_1(t,t')=
-\int_{-\infty}^{t'} dt_1\int_{-\infty}^{t_1} dt_2
f(t_1,t_2)F(t,t',t_1,t_2),
\nonumber\\&&
Q_2(t,t')=
-\int_{t'}^{t} dt_2\int_{t'}^{t_2} dt_1
f(t_1,t_2)
F(t,t_2,t_1,t'),
\nonumber\\&&
Q_3(t,t')=
-\int_{-\infty}^{t'} dt_1\int_{-\infty}^{t} dt_2
f(t_1,t_2)
F(t_1,t',t,t_2).
\end{eqnarray}
Here we denoted 
$f(t_1,t_2)={\cal E}_2(t_1){\cal E}_2(t_2)e^{i\omega_2(t_1-t_2)}$, 
and
\begin{eqnarray}\label{F}
F(t,t',t_1,t_2)=&&
\langle \Omega_0|\hat{T}e^{-iH(t-t')}\hat{T}^{\dag}e^{-iH(t'-t_1)}
\hat{T}e^{-iH(t_1-t_2)}\hat{T}^{\dag}|\Omega_0\rangle
\nonumber\\
=&&
\sum_{{\bf pq}{\bf k}'{\bf k}\lambda s \sigma' \sigma}
A_{{\bf pq}{\bf k}'{\bf k}}^{\lambda s \sigma' \sigma}
e^{-i(\varepsilon_p-\varepsilon_d)(t-t')
-i(\varepsilon_k-\varepsilon_{k'})(t'-t_1)
-i(\varepsilon_k-\varepsilon_d)(t_1-t_2)},
\end{eqnarray}
\begin{equation}
\label{relations1}
A_{{\bf pq}{\bf k}'{\bf k}}^{\lambda s \sigma' \sigma}=
\langle\Omega_0|d_{\lambda}^{\dag}c_{{\bf p}\lambda}
c_{{\bf q} s}^{\dag}d_s
d_{\sigma'}^{\dag}c_{{\bf k}'\sigma'}
c_{{\bf k}\sigma}^{\dag}d_{\sigma}|\Omega_0\rangle
=
\delta_{\lambda \sigma}\delta_{s\sigma'}
n_{\sigma}(1-n_p)
[\delta_{{\bf pk}}\delta_{{\bf qk}'}n_q
+\delta_{\sigma \sigma'}\delta_{{\bf pq}}\delta_{{\bf kk}'}(1-n_k)],
\end{equation}
with 
$n_{\sigma}=\langle \Omega_0|d_{\sigma}^{\dag}d_{\sigma}|\Omega_0\rangle$ 
and $n_k=\langle \Omega_0|c_{{\bf k}\sigma}^{\dag}c_{{\bf k}\sigma}
|\Omega_0\rangle$ (impurity occupation number is 
$n_d=\sum_{\sigma}n_\sigma=1$ here).
For monochromatic optical fields, ${\cal E}_i(t)={\cal E}_i$, the
time integrals can be explicitly evaluated. After a lengthy but
straightforward calculation, the third-order polarization
(\ref{PP3}) takes the form
$\tilde{P}^{(3)}=\tilde{P}^{(3)}_0+\tilde{P}^{(3)}_K$ 
with   

\begin{eqnarray}
\tilde{P}_0^{(3)}=&&
\mu^4{\cal E}_1{\cal E}_2^2\sum_{\bf pq}
\frac{(1-n_p)}{\varepsilon_p-\varepsilon_d-\omega_1}
\Biggl[
\frac{2}{(\varepsilon_p-\varepsilon_q)(\varepsilon_p-E_d)}
-\frac{1}{(\varepsilon_p-\varepsilon_d-\omega_1)(\varepsilon_q-E_d)}
\Biggr],\label{pol-free}
\\
\tilde{P}_{K}^{(3)}=&&
(N-1)\mu^4 {\cal E}_1{\cal E}_2^2\sum_{\bf pq}
\frac{(1-n_p)n_q}{\varepsilon_p-\varepsilon_d-\omega_1}
\Biggl[
\frac{2}{(\varepsilon_p-\varepsilon_q)(\varepsilon_p-E_d)}
-\frac{1}{(\varepsilon_p-\varepsilon_d-\omega_1)(\varepsilon_q-E_d)}
\Biggr],\label{pol-kondo}
\end{eqnarray}
where $N$ is the impurity level degeneracy.
Here we introduced the effective impurity level
$E_d=\varepsilon_d+\omega_2$. 
The first term, $\tilde{P}_0^{(3)}$, is the usual third-order
polarization for {\em spinless} ($N=1$)
electrons\cite{mukamel-book}.  The second
term, $\tilde{P}_{K}^{(3)}$, originates from the
suppression, due to the Hubbard repulsion $U$,
of the contributions from   
doubly-occupied impurity states. 
As indicated by the prefactor $(N-1)$,  
it comes from the additional intermediate states
that are absent in the spinless case  [see Fig 1(b)].

Consider the first term in Eq. (\ref{pol-kondo}).
The restriction of the sum over ${\bf q}$ to states {\em below} the
Fermi level results in a logarithmic divergence
in the absorption  coefficient, 
$\alpha\propto {\rm Im}\tilde{P}$, 
at the absorption threshold, $\omega_1=-\varepsilon_d$:
\begin{equation}
\label{alpha3}
{\rm Im}\tilde{P}_K^{(3)}=(N-1)p_0\theta(\omega_1+\varepsilon_d)
\frac{2\Delta}{\pi\delta\omega} 
\ln \biggl|\frac{D}{\omega_1+\varepsilon_d}\biggr|,
\end{equation}
where $p_0=\pi{\cal E}_1\mu^2g$, 
$\delta\omega=\omega_1-\omega_2$ is the pump-probe detuning, and
$\Delta=\pi g \mu^2 {\cal E}_2^2$
is the energy width characterizing the pump intensity;
$D$ and $g$ are the bandwidth and the density of states (per
spin) at the Fermi level, respectively.  
Recalling that the linear absorption is determined by 
${\rm Im}\tilde{P}^{(1)}=p_0 \theta(\omega_1+\varepsilon_d)$,
we  see that it differs from Eq.\ (\ref{alpha3}) by  a factor
$\frac{2\Delta}{\pi\delta\omega}
\ln \bigl|\frac{D}{\omega_1+\varepsilon_d}\bigr|$
(setting for simplicity $N=2$).
In other words,  ${\rm Im}\tilde{P}^{(1)}$ and 
${\rm Im}\tilde{P}^{(3)}_K$ become comparable when
\begin{equation}
\label{break}
\omega_1+\varepsilon_d\equiv\delta\omega+E_d
\sim D\exp\biggl(-\frac{\pi\delta\omega}{2\Delta}\biggr).
\end{equation}
We see that the   perturbative expansion 
of the nonlinear optical polarization in terms  of the optical fields
{\em breaks down} even for weak pump 
intensities (i.e.,  small  $\Delta$). The  above condition
of its validity depends critically on the detuning of the pump
frequency from the Fermi level. 
For off-resonant pump, such that the effective impurity level
lies below the Fermi level, $|E_d|=|\varepsilon_d|-\omega_2\gg \Delta$,
the relation  (\ref{break}) can be written as
$\delta\omega+E_d\sim  T_K$ with
\begin{equation}
\label{TK}
T_K=De^{\pi E_d/2\Delta}=
D\exp\biggl[-\frac{|\varepsilon_d|-\omega_2}{2g\mu^2{\cal E}_2^2}\biggr].
\end{equation}
This new energy scale can be associated with the Kondo 
temperature---an energy scale known to emerge from a spin-flip 
scattering of a FS electron by a magnetic
impurity\cite{hewson}. 
Remarkably, in our case, the Kondo 
temperature can be {\em tuned}  by varying the frequency and intensity
of the pump. In fact, the logarithmic divergence in
Eq. (\ref{alpha3}) is an indication of an {\em optically-induced}
Kondo effect. 

Let us now turn to the second term in Eq.\ (\ref{pol-kondo}). 
In fact, it represents the lowest order in the expansion of the 
linear polarization with impurity level shifted by 
$\delta\varepsilon=(N-1)\mu^2{\cal E}_2^2\sum_{\bf q}
\frac{n_q}{\varepsilon_q-E_d}$,
\begin{equation}
\label{pol-linear}
\tilde{P}^{(1)}=\mu^2 {\cal E}_1\sum_{\bf p}
\frac{(1-n_p)}{\varepsilon_p-\varepsilon_d+\delta\varepsilon-\omega_1}.
\end{equation}
The origin of $\delta\varepsilon$ can
be understood by  observing that, for {\em monochromatic} 
pump, the coupling between the 
FS and the impurity can be described by a {\em time-independent}
Anderson Hamiltonian $H_A$ with effective impurity level
$E_d=\varepsilon_d+\omega_2$ and hybridization parameter $V=\mu{\cal E}_2$. 
By virtue of this analogy, $\delta\varepsilon$ is the
perturbative solution of the following equation for the self-energy 
part:
\begin{eqnarray}
\label{eps0}
E_0=\Sigma(E_0)
\equiv (N-1)\mu^2{\cal E}_2^2\sum_{\bf q}
\frac{n_q}{\varepsilon_q-E_d+E_0}
\simeq (N-1)\frac{\Delta}{\pi}\ln \frac{E_d-E_0}{D},
\end{eqnarray}
which determines the renormalization of the effective impurity energy,
$E_d$, to $\tilde{E}_d=E_d-E_0$ \cite{hewson}.
Indeed, to the first order in the optical field, Eq. (\ref{eps0}) yields
$E_0=\delta\varepsilon$  after omitting
$E_0$ in the rhs.

The logarithmic divergence (\ref{alpha3}) indicates that near the
absorption threshold, a nonperturbative treatment is necessary.
Recall that the attractive interaction
$v_0$ between a localized hole and FS electrons also leads to a
logarithmically diverging correction (in the lowest order in $v_0$) 
even in the linear absorption:  
$\delta \tilde{P}^{(1)}\sim \tilde{P}^{(1)}gv_0
\ln[D/(\omega_1+\varepsilon_d)]$.
In the nonperturbative regime, 
$\delta \tilde{P}^{(1)}\sim \tilde{P}^{(1)}$,
this correction evolves into the Fermi edge
singularity\cite{mahan-book}. The question is how the Kondo 
correction (\ref{alpha3}) will evolve in the nonperturbative regime. 
We first discuss qualitatively our results and defer the details to the
end of the paper.

It can be seen from the expression (\ref{TK}) for $T_K$ that there
is a well-defined critical pump intensity,
$\Delta_c\equiv \pi g\mu^2{\cal E}_{2c}^2=
\frac{\pi}{2}(|\varepsilon_d| -\omega_2)$.
The shape of the nonlinear absorption spectrum  will depend sharply
on the ratio between $\Delta$ and $\Delta_c$. For strong pump, 
$\Delta >\Delta_c$, the Kondo correction (\ref{alpha3}) will develop
into a broad peak with width $\Delta$ and height $p_0$. This is
illustrated in Fig.\ 2(a).

Much more delicate is the case $\Delta\ll \Delta_c$,
which is analogous to the Kondo limit. 
The Kondo scale $T_K$ is then much smaller than $\Delta$,
which is the case for well-below-resonance pump excitation, 
$|\varepsilon_d|-\omega_2\gg \Delta$. The impurity density of states
in the Kondo limit is known\cite{hewson} 
to have two peaks well separated in energy by 
$|E_d|=|\varepsilon_d|-\omega_2 \gg\Delta$ ($E_d$ is the effective level
position). As a result, in the presence of the pump, the system sustains
excitations originating from the beats between these peaks.
These excitations can assist the absorption of a probe
photon. The corresponding condition for the probe frequency reads
$|E_d|+\omega_1\simeq |\varepsilon_d|$, or
$\omega_1\simeq\omega_2$. Thus, in the Kondo limit, the
absorption spectrum exhibits a narrow peak below the linear
absorption onset. This is illustrated in Fig.\ 2(b).

To calculate the shape of the below-threshold absorption peak, we
adopt the large $N$ variational 
wave-function method by following the approach of \cite{gunn83}.  
For monochromatic optical fields,
the polarization (\ref{PP}) can be written as
$\tilde{P}=-\mu^2{\cal E}_1[G^{<}(E_0-\delta\omega)
+G^{>}(E_0+\delta\omega)]$, where
$G^{<}(\varepsilon)=
\langle \Omega|T^{\dag}(\varepsilon -H_A)^{-1}T|\Omega\rangle$
[$G^{>}(\varepsilon)$ is similar but with $T\longleftrightarrow T^{\dag}$].
In the leading order in $N^{-1}$, $|\Omega\rangle$ is given by 
$|\Omega\rangle= A\bigl(|0\rangle 
+\sum_{\bf q}n_qa_q|{\bf q},1\rangle\bigr)$, where 
$|{\bf q},1\rangle=
N^{-1/2}\sum_{\sigma}d_{\sigma}^{\dag}c_{{\bf q}\sigma}|0\rangle$
($|0\rangle$ stands for the full FS). The coefficients $A$ and $a_k$ are
found by minimizing $H_A$ in this basis; one then obtains, e.g.,
$A^2=1-n_d$, where  $n_d=(1+\pi \tilde{E}_d/N\Delta)^{-1}$
is the impurity occupation\cite{gunn83,hewson}
($N\Delta$ is finite in the large $N$ limit). The relevant Green
function is obtained as
\begin{equation}
\label{green2}
G^{<}(\varepsilon)
=\frac{\pi}{\Delta}\Biggl[\Sigma(\varepsilon)
+
\frac{|\Sigma(\varepsilon)|^2}
{\varepsilon-\Sigma(\varepsilon)}
\Biggr].
\end{equation}
Since $\Sigma(E_0)=E_0$ [see Eq. (\ref{eps0})], for 
$\varepsilon=E_0-\delta\omega$ the second 
term has a pole at $\delta\omega=0$ which gives rise to a resonance.
The $N^{-1}$ correction gives a finite resonance width $\Delta$. 
Using that the residue at the pole is 
$[\partial\Sigma(E_0)/\partial E_0-1]^{-1}=n_d-1$ \cite{hewson}, 
we finally obtain

\begin{eqnarray}
\label{kondo-abs}
{\rm Im}\tilde{P}_K= 
\frac{p_0E_0^2(1-n_d)^2}
{\delta\omega^2+\Delta^2}
\sim 
\biggl(\frac{\pi E_d T_K}{N\Delta}\biggr)^2
\! \frac{p_0}{\delta\omega^2+\Delta^2}.
\end{eqnarray}
For the last estimate, we used that, in the Kondo limit 
($\Delta\ll \Delta_c$), $1-n_d\simeq \pi T_K/N\Delta$ 
and $E_0\simeq E_d$. Then the rhs of (\ref{kondo-abs}) describes 
the narrow below-threshold peak [see Fig.\ 2(b)]. 
In the Kondo limit, the factor $(1-n_d)^2$ has the physical meaning of a
product of populations of electrons in the narrow peak of the impurity
spectral function (Kondo resonance) and ``holes'' in the wide peak
(centered at $\varepsilon_d$ below the Fermi level). 
Note, however that
the above calculation was not restricted to the Kondo limit. For 
$\Delta\gtrsim \Delta_c$ (mixed-valence regime), we have 
$1-n_d\sim 1$ and $E_0 \sim N\Delta$. Then Eq. (\ref{kondo-abs})
reproduces the absorption peak in Fig.\ 2(a).

Note that, although we considered here, for simplicity, the limit of 
singly occupied impurity level in the ground state, 
the Kondo-absorption can take place even if the impurity 
is {\em doubly} occupied. 
Indeed, after the probe excites an impurity electron,
the spin-flip scattering of FS electrons with the remaining 
impurity electron will lead to the Kondo resonance in
the final state of the transition. In this case, however, the Kondo
effect should show up in the fifth-order polarization.

A feasible system in which the proposed effect might be observed is,
e.g., GaAs/AlGaAs superlattice delta-doped with Si donors located in 
the barrier. The role of impurity in this
system is played by a shallow acceptor, e.g., Be. Molecular-beam
epitaxy growth 
technology allows one to vary the quantum well width and to place
acceptors right in the middle of each quantum well \cite{acceptor}.
In quantum wells, the valence band is only doubly degenerate with
respect to the total angular momentum J. Thus, such a system
emulates the large U limit considered here. The  dipole matrix
element for acceptor to conduction band transitions can be estimated
as $\mu\sim\mu_0 a$, where $\mu_0$ is the interband matrix element
and $a$ is the size of the acceptor wave function. 
For typical excitation intensities\cite{reviews}, the parameter
$\Delta$ ranges on the meV scale resulting in 
$T_K\sim \Delta$ for the pump detuning of several meV.

In conclusion, let us discuss the effect of a finite duration of the pump
pulse, $\tau$. Our result for $\chi^{(3)}$ remains unchanged if $\tau$
is longer than $\hbar/T_K$. 
If $\tau < \hbar/T_K$, then $\tau$ will serve as a cutoff 
of  the logarithmic divergence in (\ref{alpha3}), and the 
Kondo correction will depend on the parameters of the pump
${\cal E}_2$ and $\tau$ as follows: 
$ {\rm Im}\tilde{P}_K^{(3)}\propto {\cal E}_2^2\ln(D\tau/\hbar)$.
In the non-perturbative regime, our basic assumption was that, for
monochromatic pump, the system maps onto the {\em ground} state of the
Anderson Hamiltonian. Our results apply if the pump is
turned on slowly on a time scale longer than
$\hbar/T_K$. For shorter 
pulse duration, the build up of the optically-induced Kondo effect
will depend on the dephasing of FS excitations\cite{FES}.
The role of interactions between FS and impurity electrons
in the presence of hybridization was addressed in 
\cite{perakis93}. An avenue for future studies would be the
interplay between the Kondo-absorption and the Fermi edge singularity.
Note finally that the effect of irradiation on the Kondo transport in
quantum dots was investigated in \cite{glazman99,dots}.

This work was supported by NSF grants ECS-9703453 (Vanderbilt)
and DMR 9732820 (Utah), and Petroleum Research Fund grant 
ACS-PRF \#34302-AC6 (Utah).


\begin{figure}
\caption{
Intermediate processes contributing to $\chi^{(3)}$. 
(a) Intermediate state with doubly-occupied impurity. 
(b) Large $U$ limit: {\em Two} transition channels are available from
states {\em below} the FS to the {\em empty} impurity, but only {\em one} 
channel from the {\em singly-occupied} impurity to states {\em above} the FS. 
}

\end{figure}

\begin{figure}
\caption{Schematic plot of the nonlinear absorption spectra vs probe
frequency. (a) Mixed-valence regime: pump-probe spectrum for strong pump
intensity (thick line) compared with $\chi^{(3)}$ approximation
(\ref{alpha3}) (dashed line) and the linear absorption spectrum (thin line).
(b) Kondo limit: the pump-probe spectrum has a narrow
peak below the linear absorption threshold.}
\end{figure}


\clearpage
\begin{center}
\epsfxsize=6.0in
\epsffile{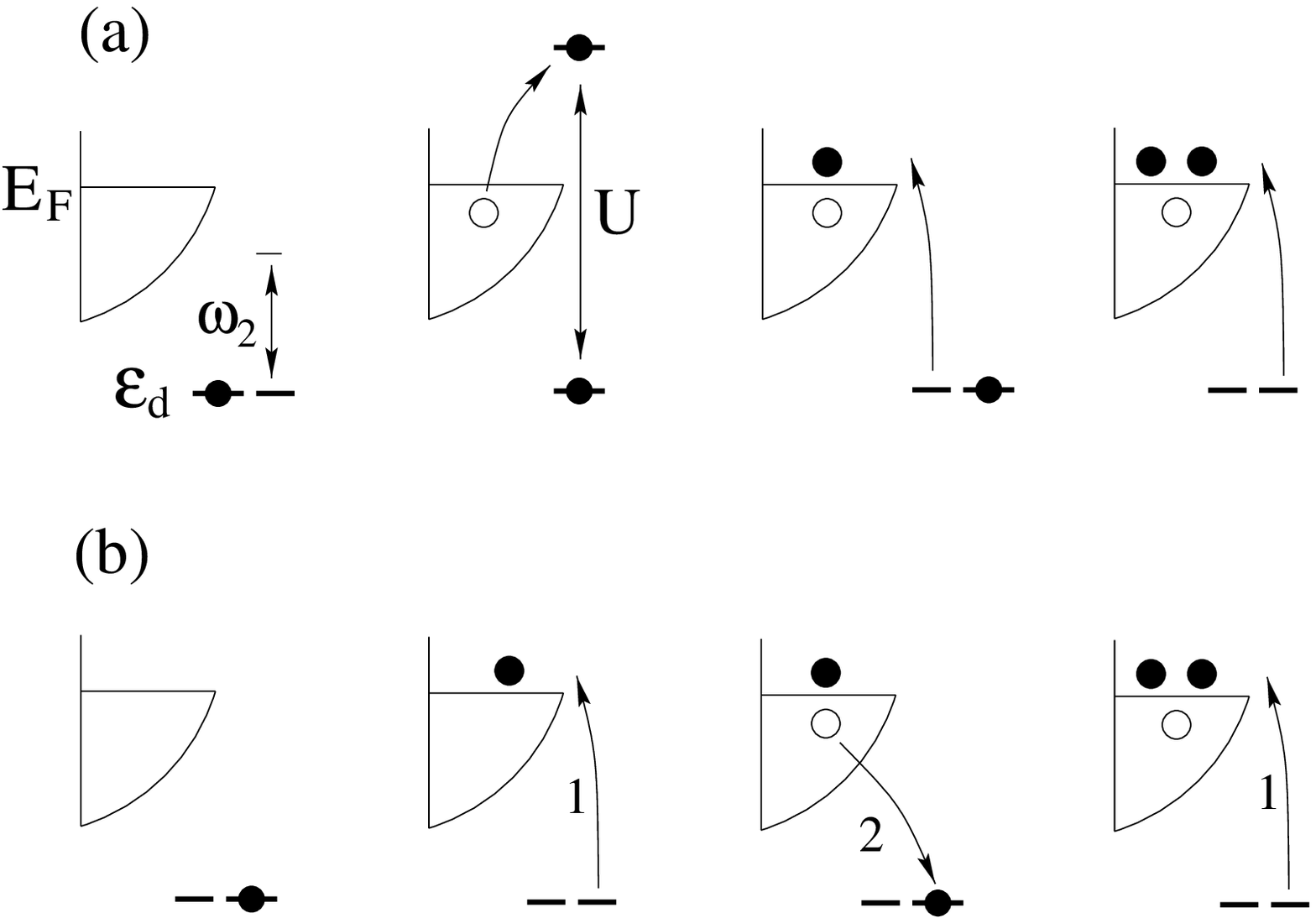}
\vspace{80mm}
\centerline{Fig.\ 1}
\clearpage
\epsfxsize=5.0in
\epsffile{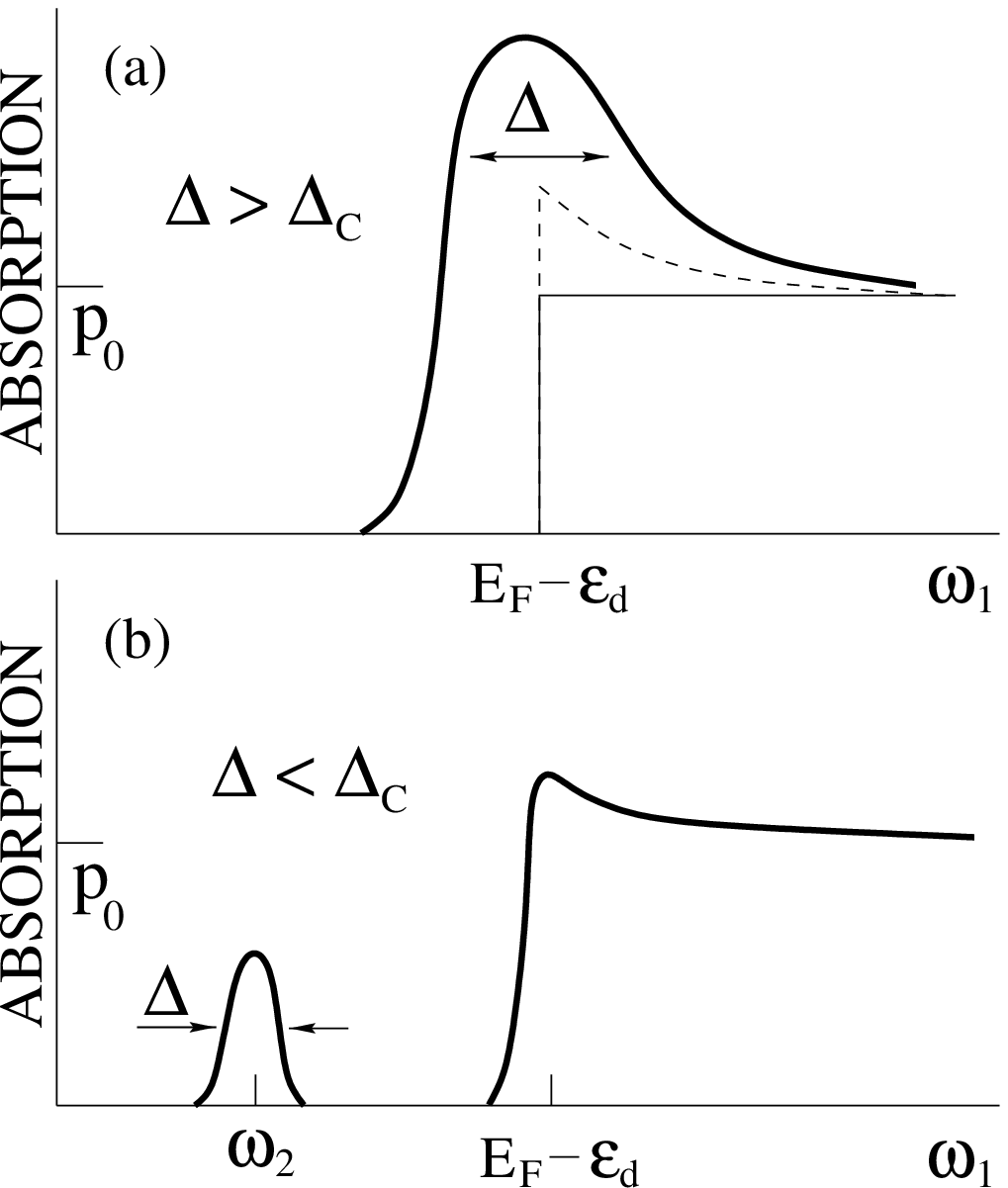}
\end{center}
\vspace{40mm}
\centerline{Fig.\ 2}


\begin{references}
\bibitem{mahan-book} 
G.\ D.\ Mahan, 
{\em Many-Particle Physics}\ (Plenum,\ 1990).

\bibitem{reviews} 
D. S. Chemla, in 
{\em Nonlinear Optics in Semiconductors}, edited
by R.\ K.\ Willardson and A.\ C.\ Beers\ (Academic\ Press,\ 1999); 
V.\ M.\ Axt\ and\ S.\ Mukamel,\ Rev.\ Mod.\ Phys.\ {\bf 70},\ 145\ (1998). 

\bibitem{bre95} 
I.\ Brener,
W. H. Knox, and W. Schaefer,
Phys. Rev. B {\bf 51}, 2005 (1995);
I.\ E.\ Perakis\ {\em et\ al.},
J.\ Opt.\ Soc.\ Am.\ B\ {\bf 13},\ 1313\ (1996).

\bibitem{portengen}T. Portengen, Th. Ostreich, and L. J. Sham, 
Phys. Rev. Lett. {\bf 76}, 3384 (1996). 

\bibitem{awschalom}
J. M. Kikkawa and D. D. Awschalom, 
Phys.\ Rev.\ Lett.\ {\bf 74},\ 80\ 4313 (1998).

\bibitem{hall}N.\ A.\ Fromer\ {\em et al.},
Phys.\ Rev.\ Lett.\ {\bf 83},\ 4646\ (1999). 

\bibitem{dodge99}S.\ Dodge\ {\em et\ al.},
Phys.\ Rev.\ Lett.\ {\bf 83},\ 4650\ (1999).

\bibitem{FES}
T.\ V.\ Shahbazyan\ {\em et\ al.},
Phys.\ Rev.\ Lett.\ {\bf 84},\ 2006\ (2000);
N.\ Primozich\ {\em et al.},
Phys.\ Rev.\ B\ {\bf 61},\ 2041\ (2000).

\bibitem{mukamel-book}S.\ Mukamel, 
{\em Principles of Nonlinear Optical Spectroscopy}
(Oxford University Press, 1995).


\bibitem{hewson}A.\ C.\ Hewson, 
{\em The Kondo Problem to Heavy Fermions}
(Cambridge University Press, 1993).


\bibitem{gunn83}O.\ Gunnarson and K.\ Sch\"{o}nhammer,
Phys.\ Rev.\ B\ {\bf 28},\ 4315\ (1983).

\bibitem{acceptor}P.\ O.\ Holtz\ {\em et\ al.}, 
Phys.\ Rev.\ B\ {\bf 47},\ 15 675\ (1993);
I. V. Kukushkin\ {\em et\ al.},
Phys.\ Rev.\ B\ {\bf 40},\ 7788\ (1989).

\bibitem{perakis93}I. E. Perakis,
C. M. Varma, and A. E. Ruckenstein, 
Phys.\ Rev.\ Lett.\ {\bf 70},\ 3467\ (1993); 
I.\ E. Perakis and C.\ M. Varma, Phys.\ Rev.\ B\ {\bf 49},\ 9041\ (1994).


\bibitem{glazman99}A. Kaminski,
Yu. Nazarov, and L. I. Glazman,
Phys.\ Rev.\ Lett. {\bf 83},\ 384\ (1999).

\bibitem{dots}M.\ H.\ Hettler and H.\ Sch\"{o}ller, 
Phys.\ Rev.\ Lett.\ {\bf 74},\ 4907\ (1995); 
T.\ K.\ Ng, {\em ibid.}\ {\bf 76},\ 487\ (1996); 
R.\ Lopes\ {\em et\ al.},\ {\em ibid.}\ {\bf 81},\ 4688\ (1998);
Y.\ Goldin and Y.\ Avishai,  {\em ibid.}\ {\bf 81},\ 5394\ (1998);
P. Nordlander\ {\em et\ al.},\ {\em ibid.}\ {\bf 83},\ 808\ (1999).

\end{references}
\end{document}